\documentclass[]{JHEP3}

\usepackage{amsfonts,amsbsy,amsmath,amssymb}

\usepackage[mathscr]{eucal}
\allowdisplaybreaks[4]

\title{New Terms for the Compact Form of Electroweak Chiral Lagrangian}

\author{Hong-Hao Zhang, Wen-Bin Yan, and J. K. Parry\\
Center for High Energy Physics $\&$ Department of Physics, Tsinghua
University,\\ Beijing 100084, China\\
E-mail: \email{zhanghonghao@tsinghua.org.cn},
\email{wenbin.yan@gmail.com}, \email{jkparry@tsinghua.edu.cn}}

\author{Xue-Song Li\\
Science College, Hunan Agricultural University, Changsha 410128,
China\\
E-mail: \email{lixuesong@tsinghua.org.cn}}

\abstract{The compact form of the electroweak chiral Lagrangian is a
reformulation of its original form and is expressed in terms of
chiral rotated electroweak gauge fields, which is crucial for
relating the information of underlying theories to the coefficients
of the low-energy effective Lagrangian. However the compact form
obtained in previous works is not complete. In this letter we add
several new chiral invariant terms to it and discuss the
contributions of these terms to the original electroweak chiral
Lagrangian.}

\preprint{TUHEP-TH-07156} \keywords{Electroweak Chiral Lagrangian,
Beyond Standard Model}

\begin{document}

So far the postulated Higgs particle of the standard model has not
been observed in experiments. We do not know what the electroweak
symmetry breaking mechanism in nature is. The electroweak chiral
Lagrangian is a general low-energy description for electroweak
symmetry breaking patterns
\cite{Appelquist:1980vg,Longhitano:1980iz}, especially for those
strong dynamical electroweak symmetry breaking mechanisms
\cite{Hill:2002ap}. All the coefficients of the electroweak chiral
Lagrangian, in principle, can be fixed by experiments
\cite{Donoghue:1988ed,He:1996nm,He:1997kn,Alam:1997nk,He:2002qi,Zhang:2003it,
He:2004ug,Chivukula:2005ji,Han:2005pu,Boos:1997gw}. Due to the
nonperturbative property of the possible strong dynamics, it is
difficult to relate the measured coefficients of the electroweak
chiral Lagrangian to underlying theories. Recently, the series of
work of Ref.
\cite{Wang:1999cp,Wang:1999xh,Wang:2000mg,Wang:2000mu,Yang:2002he,
Wang:2002rb,Wang:2005cm,Ma:2003uv} successfully produced the
predictions for the coefficients of the chiral Lagrangian from the
underlying theory of QCD, which lights up the hope of building up
the relationship between the coefficients of the electroweak chiral
Lagrangian and underlying strong dynamical models. As seen in Ref.
\cite{Wang:1999cp}, a crucial first step of the derivation of the
chiral Lagrangian from QCD is to reformulate the original chiral
Lagrangian in terms of chiral rotated external source fields. For
the case of the electroweak chiral Lagrangian, we need to similarly
reformulate it in terms of chiral rotated electroweak gauge fields
in order to deduce the information of underlying theories. An
attempt of this line of thought is the work of Ref.
\cite{Wang:2001xz}, which tries to give a so-called compact form of
the electroweak chiral Lagrangian in terms of chiral rotated
electroweak gauge fields. However, there are still several relevant
terms not included in the reformulation of Ref. \cite{Wang:2001xz}.
In this letter, we shall present 3 new terms for this reformulation
and give their contributions to the original electroweak chiral
Lagrangian.

We refer the interested reader to Ref. \cite{Wang:2001xz} for all
the details on the compact form of the electroweak chiral Lagrangian
and the relations between the compact form and the original form.
Besides the inner-product terms
$c_2g_2^4V_\mu^{a'}V_\nu^{a'}V^{b'\mu}V^{b'\nu}$, $c_8g_2^2(d_\mu
V_\nu^{a'}-d_\nu V_\mu^{a'})V^{a'\mu}A^\nu$, and $c_9g_2^2(d^\mu
V_\mu^{a'})V_\nu^{a'}A^\nu$ already included in the compact form of
Ref. \cite{Wang:2001xz}, there should be 3 new cross-product terms
which are given by,
\begin{eqnarray}
\Delta\mathcal{L}&=&c'_2g_2^4(\epsilon^{a'b'}V_\mu^{a'}V_\nu^{b'})^2
+c'_8g_2^2\epsilon^{a'b'}(d_\mu V_\nu^{a'}-d_\nu
V_\mu^{a'})V^{b'\mu}A^\nu+c'_9g_2^2\epsilon^{a'b'}(d^\mu
V_\mu^{a'})V_\nu^{b'}A^\nu\;,\label{new-terms}
\end{eqnarray}
where $a'$, $b'$ run from 1 to 2, and
$\epsilon^{12}=-\epsilon^{21}=1$. Here and henceforth $V^{a'\mu}$
and $A^\mu$ are short for the chiral rotated gauge fields
$V_\xi^{a'\mu}$ and $A_\xi^\mu$ in Ref. \cite{Wang:2001xz}
respectively. There are also 3 cross-product terms corresponding to
the $c_{15,17,25}$-terms in that paper, but they vanish. Let us
consider Eq. \eqref{new-terms} term by term. From the definitions of
Ref. \cite{Wang:2001xz}, it is straightforward to obtain the
relation between the first term of Eq. \eqref{new-terms} and the
ordinary terms of the electroweak chiral Lagrangian as follows,
\begin{eqnarray}
c'_2g_2^4(\epsilon^{a'b'}V_\mu^{a'}V_\nu^{b'})^2
&=&-4c'_2\bigg[[{\rm tr}(X_\mu X_\nu)]^2-[{\rm tr}(X_\mu
X^\mu)]^2-{\rm tr}(X_\mu X_\nu){\rm tr}(\tau^3X^\mu){\rm
tr}(\tau^3X^\nu)\nonumber\\
&&+{\rm tr}(X_\mu X^\mu)[{\rm
tr}(\tau^3X_\nu)]^2\bigg]\;,\label{new-1}
\end{eqnarray}
where $X_\mu\equiv U^\dag (D_\mu U)$. And the last two terms of Eq.
\eqref{new-terms} are respectively given by,
\begin{eqnarray}
&&c'_8g_2^2\epsilon^{a'b'}(d_\mu V_\nu^{a'}-d_\nu
V_\mu^{a'})V^{b'\mu}A^\nu\nonumber\\
&&=c'_8\bigg[-\frac{1}{2}\mathrm{tr}(\tau^3X^\nu)
\big[\mathrm{tr}(\tau^3X^\mu)\mathrm{tr}(X_\mu
X_\nu)-\mathrm{tr}(\tau^3X_\nu)\mathrm{tr}(X_\mu X^\mu)\big]
+i\mathrm{tr}(\overline{W}_{\mu\nu}X^\mu X^\nu)\nonumber\\
&&~~-\frac{i}{2}\mathrm{tr}(\tau^3\overline{W}_{\mu\nu})\mathrm{tr}(\tau^3X^\mu
X^\nu)\bigg]\;,\label{new-2}
\end{eqnarray}
with $\overline{W}_{\mu\nu}\equiv U^\dag
g_2\frac{\tau^a}{2}W_{\mu\nu}^aU$, and
\begin{eqnarray}
c'_9g_2^2\epsilon^{a'b'}(d^\mu
V_\mu^{a'})V_\nu^{b'}A^\nu&=&\frac{1}{2}(1-4\beta_1)c'_9\bigg[
\mathrm{tr}(\tau^3X^\mu)\mathrm{tr}(\tau^3X^\nu)\mathrm{tr}(X_\mu
X_\nu)\nonumber\\
&&-[\mathrm{tr}(\tau^3X_\mu)\mathrm{tr}(\tau^3X_\nu)]^2\bigg]\;.\label{new-3}
\end{eqnarray}
From Eqs. \eqref{new-1}, \eqref{new-2} and \eqref{new-3}, we obtain
the contributions of these 3 new terms to the original electroweak
chiral Lagrangian as follows,
\begin{eqnarray}
&&\Delta\alpha_3=\frac{c'_8}{2}\;,\qquad
\Delta\alpha_4=-4c'_2\;,\qquad \Delta\alpha_5=4c'_2\;,\nonumber\\
&&\Delta\alpha_6=4c'_2-\frac{c'_8}{2}+\frac{1}{2}(1-4\beta_1)c'_9\;,\qquad
\Delta\alpha_7=-4c'_2+\frac{c'_8}{2}\;,\nonumber\\
&&\Delta\alpha_9=-\frac{c'_8}{2}\;,\qquad
\Delta\alpha_{10}=(1-4\beta_1)c'_9\;.
\end{eqnarray}
The coefficients $c'_{2,8,9}$ and $c_i$ ($i=1,2,\ldots,25$) in this
compact form of the electroweak chiral Lagrangian are determined by
the underlying ultraviolet theories. For example, if we take the
one-doublet technicolor model as the underlying theory, it can be
shown that these 3 new coefficients $c'_{2,8,9}$ are all
non-vanishing, and full details will be presented in forthcoming
publications \cite{Zhang:2007zi}.

In summary, we have provided 3 new terms to the compact form of the
electroweak chiral Lagrangian introduced in Ref. \cite{Wang:2001xz}.
These additional terms were not considered in the previous work. In
this letter we have related these new terms to the original
electroweak chiral Lagrangian, which will be crucial in forthcoming
studies of strong dynamical models.

%%%%%%%%%%%%%%%%%%%%%%%%%%%%%%%%%%%%%%%%%%%%%%%%%%%%%%
\section*{Acknowledgments}
We are indebted to Qing Wang for all our knowledge about the chiral
Lagrangian and his helps and supports for this work. This work is
supported in part by the National Natural Science Foundation of
China.
%%%%%%%%%%%%%%%%%%%%%%%%%%%%%%%%5

\end{document}